\documentclass[pra,twocolumn,showpacs,preprintnumbers,amsmath,amssymb,superscriptaddress]{revtex4}

\usepackage{graphicx}
\usepackage{dcolumn}
\usepackage{bm}
\usepackage{amsmath}
\usepackage{amssymb}
\usepackage{amsthm}

\begin{document}

\title{Optical characterization and selective addressing of the resonant modes of a micropillar cavity with a white light beam}

\author{Georgios Ctistis}\email{g.ctistis@utwente.nl}
\affiliation{Complex Photonic Systems (COPS), MESA+ Institute for
Nanotechnology, University of Twente, 7500 AE Enschede, The Netherlands}
\affiliation{Center for Nanophotonics, FOM Institute for Atomic and
Molecular Physics (AMOLF), Science Park 113, 1098 XG Amsterdam,
The Netherlands}
\author{Alex Hartsuiker}
\author{Edwin van der Pol}
\affiliation{Complex Photonic Systems (COPS), MESA+ Institute for
Nanotechnology, University of Twente, 7500 AE Enschede, The Netherlands}
\affiliation{Center for Nanophotonics, FOM Institute for Atomic and
Molecular Physics (AMOLF), Science Park 113, 1098 XG Amsterdam,
The Netherlands}%

\author{Julien Claudon}
\affiliation{CEA/INAC/SP2M, Nanophysics and Semiconductor
Laboratory, 17 rue des Martyrs, 38054 Grenoble Cedex, France}

\author{Willem L. Vos}
\affiliation{Complex Photonic Systems (COPS), MESA+ Institute for
Nanotechnology, University of Twente, 7500 AE Enschede, The Netherlands}

\author{Jean-Michel G\'erard}
\affiliation{CEA/INAC/SP2M, Nanophysics and Semiconductor
Laboratory, 17 rue des Martyrs, 38054 Grenoble Cedex, France}

\begin{abstract}
We have performed white-light reflectivity measurements on GaAs/AlAs micropillar cavities with diameters ranging from $1\ \rm{\mu m}$ up to $20\ \rm{\mu m}$. We are able to resolve the spatial field distribution of each cavity mode in real space by scanning a small-sized beam across the top facet of each micropillar. We spectrally resolve distinct transverse optical cavity modes in reflectivity. Using this procedure we can selectively address a single mode in the multimode micropillar cavity. Calculations for the coupling efficiency of a small-diameter beam to each mode are in very good agreement with our reflectivity measurements.
\end{abstract}

\maketitle

\section{Introduction}

Semiconductor cavities have attracted considerable attention in recent years due to their ability to confine light in all three dimensions, which is a key issue in solid-state cavity-quantum electrodynamic (cQED) experiments \cite{Vahala2003, Gerard2003}.
Due to its well-defined, directional radiation pattern, the micropillar geometry has been widely used over the past 20 years. Besides fundamental cQED experiments such as the demonstration of the Purcell-effect \cite{Gerard1998, Graham1999, Moreau2001, Santori2002, Varoutsis2005, Lohmeyer2006} and vacuum Rabi splitting \cite{Reithmaier2004, Khitrova2006}, 
they are also used in applications such as low-threshold vertical-cavity surface-emitting lasers (VCSELs) \cite{Huffaker1997}, all-optical switches \cite{Jewell1989, Rivera1994}, and single-mode single photon sources \cite{Moreau2001, Santori2002, Varoutsis2005, Xu2008, Heindel2010} or sources of entangled photon pairs generated by parametric polariton luminescence \cite{Bajoni2007}.

To achieve pronounced cQED effects a high quality factor $Q$ is important, since it is proportional to the photon storage time $\tau_{cav}$ in the cavity.
Furthermore, a small mode volume $V_{mode}$ is important because the coupling strength $g$ between the light field and an emitter increases with $1/\sqrt{V_{mode}}$. 
Micropillar cavities fulfill these requirements as they exhibit $Q$-factors as large as $10^5$ \cite{Reitzenstein2007} with mode volumes of the order of few $(\lambda/n)^3$. 

To understand all underlying processes in cQED experiments using micropillar cavities, it is essential to get precise information on their discrete cavity modes, \textit{i.e.}, the mode frequency and quality factor $Q$ as well as the spatial distribution of the electromagnetic field.
To get this information, previous studies performed reflectivity measurements using a narrowband laser and exciting only a single mode \cite{Jewell1989, Rivera1999}, or photoluminescence experiments with a broadband internal light source \cite{Gerard1996, Gutbrod1999, Reitzenstein2007, Constantin2002, Rivera1999, Kistner2009, Reitzenstein2009}.

A major drawback of reflectivity measurements is that several constraints apply to the coupling of an external beam into a resonant mode of a pillar cavity.
When a narrowband laser is used, frequency-matching imposes a fine spectral tuning of the source.
Furthermore, the beam-to-mode coupling also depends on the spatial overlap of their field distributions. When the position and waist of the beam is matched to a micropillar, symmetry constraints only allow a coupling of the external beam to a few modes \cite{Jewell1989, Rivera1999}.
This explains why the vast majority of experiments conducted to date use photoluminescence from an internal light source, such as an array of quantum dots, to study the cavity modes. 
Such experiments allow to measure frequency, $Q$ \cite{Gerard1996}, and far-field radiation pattern for all modes \cite{Gerard1996, Gutbrod1999}, albeit at the price of some additional losses induced by the absorption of the internal emitters. 
It is therefore interesting and important to develop novel methods enabling a systematic probing of the modes of an empty micropillar.

Here, we present a novel method for the systematic characterization of cavity modes in micropillar cavities. 
We use a broadband light source in reflection configuration allowing us to probe all modes of a cavity at once. 
Since our white-light coherent beam is tightly focused with a waist much smaller than the pillar diameter, we lift symmetry constraints that would inhibit the coupling to different modes.
As a result we are able to probe the spatial distribution of all micropillar modes. 
Furthermore, we address a single mode with our coherent beam, while being able to probe the full spectrum at once.
The proposed method leads to a direct experimental access of the modes, the modal field distribution and to the modal $Q$-factors from the outside of the micropillar cavity. 
It is thereby a fast and convenient method that opens prospects for ultrafast all-optical switching experiments on micropillars.

The outline of this paper is as follows: A brief discussion of the samples and the experimental setup is followed by the experimental results. We present data showing the strength of the setup to couple light into all modes of micropillar cavities and the ability to specifically address a single one.  
A short conclusion with a prospect to possible experiments closes the paper.

\section{Experimental}

The fabrication process of our micropillars is divided into two steps, a growth and a subsequent etching step. The growth process starts with the growth of a planar microcavity by means of molecular beam epitaxy on a GaAs(001) substrate at a temperature of $550^{\circ}\rm{C}$. During growth, 4 layers of InGaAs quantum dots with a density of $10^{10}\ \rm{cm^{-2}}$ were grown inside the GaAs $\lambda$-layer for photoluminescence measurements. 
The quantum dots are expected to have a negligible influence on the measured spectra \cite{Cargese1998}. We will use these internal light sources to cross-check our reflectivity results with photoluminescence experiments.
From the planar structure, micropillars have been etched by a reactive ion-etching (RIE) process at room temperature, resulting in micropillar cavities with diameters in the range between $1\ \rm{\mu m}$ and $20\ \rm{\mu m}$.
During the reactive ion-etching process, a thin layer of SiO$_{x}$ with thickness between $100$ and $200\ \rm{nm}$ is deposited on the sidewalls; this layer prevents oxidation of the AlAs in the Bragg stacks.
More details on the fabrication process have been reported in Ref. \cite{Gerard1996}.

\begin{figure}
\includegraphics[]{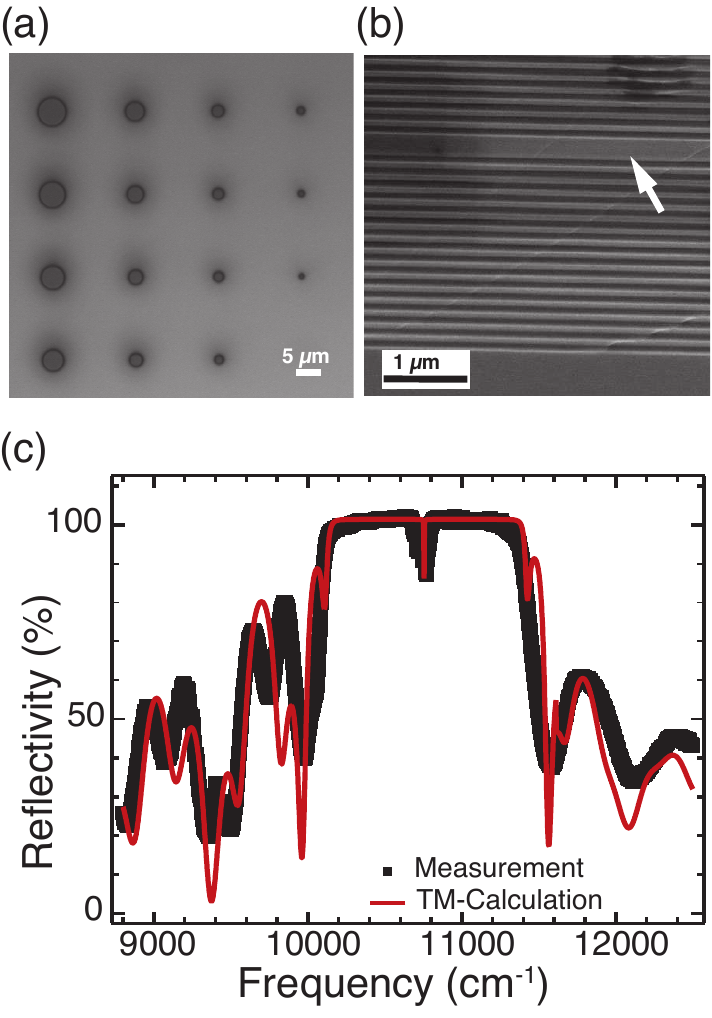}
\caption{\label{fig:fig1} (color online) (a) Top view optical microscopy image of a micropillar field with pillar diameters ranging from $6\ \mu m$ (upper left corner) to $1\ \rm{\mu m}$ diameter (lower right). (b) Scanning electron micrograph cross-section of a cleaved micropillar. The alternating shadings denote the alternating GaAs/AlAs $\lambda/4$-layers in the Bragg stacks. The arrow indicates the position of the GaAs $\lambda$-layer, having a thickness of $\rm{d=261.7}~\rm{nm}$. (c) White-light reflectivity measurement on a $20\ \rm{\mu m}$ diameter micropillar cavity (squares) and a transfer matrix calculation model (curve). The calculation fits the experimental spectrum well and thus explains all observed features. The stopband, ranging from $10100\ \rm{cm^{-1}}$ to $11400\ \rm{cm^{-1}}$ has a relative bandwidth of $12.8\ \%$ and the resonance of the cavity is at $\omega_{cav}=10755\ \rm{cm^{-1}}$.}
\end{figure}

Figure \ref{fig:fig1}(a) shows an optical microscopy image of a typical micropillar field with micropillar diameters ranging from $6\ \rm{\mu m}$ (top left corner) to $1\ \rm{\mu m}$ (bottom right), respectively. The scanning electron micrograph [Fig. \ref{fig:fig1}(b)] shows a cross-section of a cleaved micropillar. 
The alternating light- and dark-gray shadings denote the alternating GaAs and AlAs $\lambda/4$-layers of the Bragg stacks. The white arrow indicates the position of the GaAs $\lambda$-layer.  

\begin{figure}
\includegraphics[]{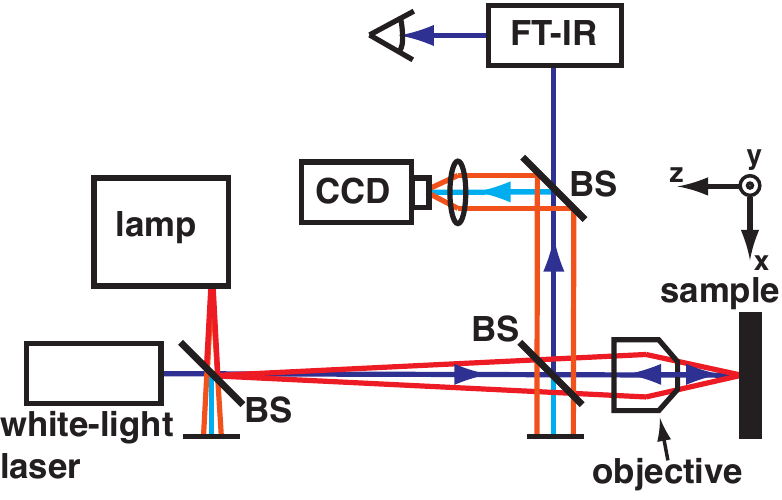}
\caption{\label{fig:fig2} (color online) Schematic of the reflectivity setup. The beam from a broadband white-light laser source is focused onto the sample using a gold-coated dispersionless reflecting microscope objective with NA 0.65. The reflected signal is then analyzed using a FT-IR spectrometer and a Si detector. Separately, light from a halogen lamp illuminates the sample and is then focused onto a CCD camera to monitor the positioning of the sample with respect to the laser beam.}
\end{figure}

We have performed room temperature reflectivity measurements using the setup shown in Fig. \ref{fig:fig2}. It consists of a white-light laser source (Fiamium, SC-450), a high-NA reflecting objective (Ealing, $\rm{NA}=0.65$) and a Fourier-transform infrared spectrometer (FTIR, BioRad, FTS-6000) equipped with a silicon photodiode.
The white-light source covers a spectral range from $4000\ \rm{cm^{-1}}$ to $22222\ \rm{cm^{-1}}$.
The white-light beam is highly collimated and can easily be focused down to its diffraction limit.
The diameter of the beam is estimated to $\rm{d}\approx1\ \rm{\mu m}$, thus nearly diffraction limited for the cavity resonance frequency.
The FTIR-setup is similar to the one used by Ref. \cite{Thijssen1999} and has a resolution of $1\ \rm{cm^{-1}}$.
To accurately position the beam on the top facet of a micropillar, the sample is mounted on an automated xyz-translation stage with a positioning accuracy of $\sim50\ \rm{nm}$. 
Furthermore, we simultaneously observed the positioning of the beam with respect to the top facet of the micropillar with an optical microscope equipped with a CCD camera.
A schematic of the setup is shown in Fig. \ref{fig:fig2}.
The reflectivity of the micropillars has been calibrated using spectra from a gold mirror. 
We observed a systematic difference in resonance frequency $\omega$ between different micropillars due to a spatial gradient in the cavity thickness across the micropillar fields as a result of the fabrication process. We have corrected for this gradient.

\section{Spectral characterization of the pillar modes}

Figure \ref{fig:fig1}(c) shows a measured reflectivity spectrum of a $20\ \rm{\mu m}$ diameter micropillar. One can clearly see the cavity resonance at $\omega_{cav}=10755\ cm^{-1}$ ($\lambda_{cav}=929.8\ \rm{nm}$) as a trough inside the stopband.
The stopband has a reflectivity close to $100\ \%$ and a relative width of $12.8\ \%$.  
A transfer-matrix calculation \cite{Born2002} is plotted into Fig. \ref{fig:fig1}(c).
The calculations were performed with 13 $\lambda/4$ pairs of GaAs/AlAs as top Bragg stack, a $261.7\ nm$ thick GaAs $\lambda$-layer, and a Bragg stack consisting of 25 $\lambda/4$ pairs of GaAs/AlAs at the bottom. The thickness of the GaAs layers is $65.4\ nm$ and the thickness of the AlAs layers is $78.3\ nm$.  
The spectral features are somewhat less pronounced in the measurement compared to the calculation, 
since the high numerical aperture of the microscope objective broadens the spectral features.
Furthermore, the calculation has been performed on a planar structure while the micropillar has a finite size in the plane. 
All in all, the calculated spectrum corresponds well to the measured one.

\begin{figure}
\includegraphics[]{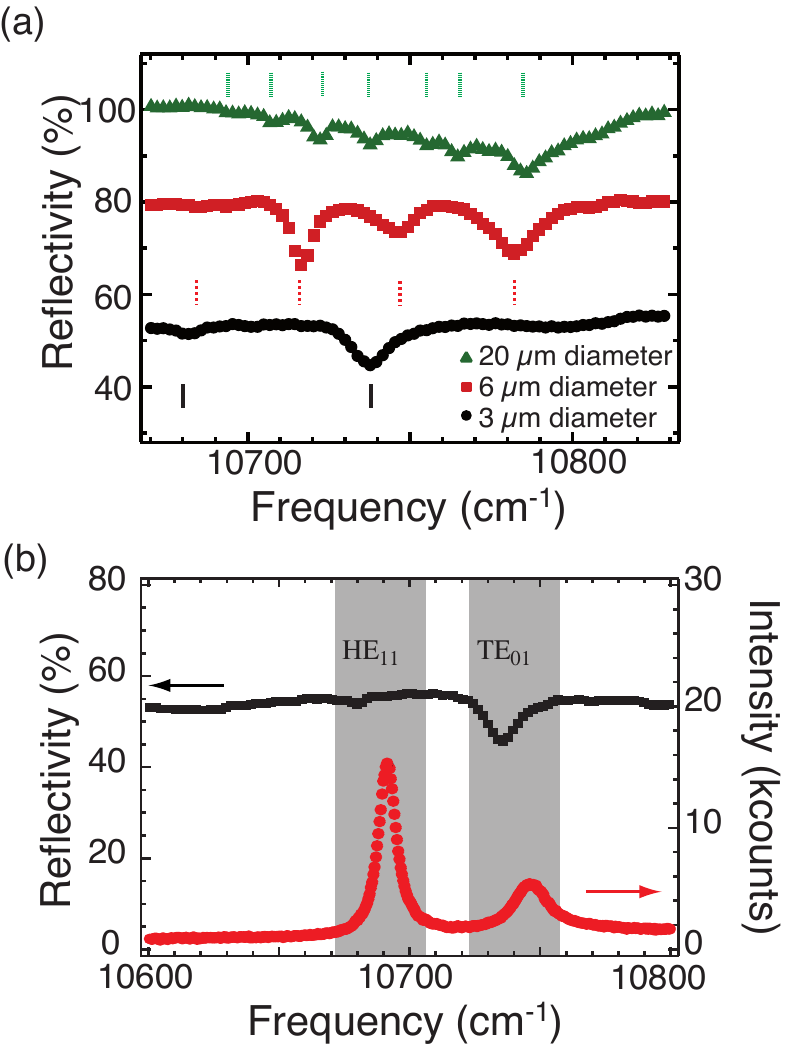}
\caption{\label{fig:fig3}(color online) (a) Detailed reflectivity spectra of the cavity resonance for three different micropillar diameters: $20\ \rm{\mu m}$ (triangles), $6\ \rm{\mu m}$ (squares), and $3\ \rm{\mu m}$ (circles), respectively. The number of observed modes increases with pillar diameter. The bars denote the frequency of each measured mode. The spectra of the $6\ \rm{\mu m}$ and $3\ \rm{\mu m}$ pillar are shifted down for clarity. (b) Comparison between the reflectivity (squares) and photoluminescence (circles) spectra of a $3\ \rm{\mu m}$ diameter pillar. The spacings between the observed modes are $55\ cm^{-1}$ (reflectivity) and $56\ cm^{-1}$ (photoluminescence), respectively.}
\end {figure}

A zoom-in into the cavity resonance of the reflectivity spectrum is shown in Fig. \ref{fig:fig3}(a) for three different micropillar diameters. 
The major feature visible in the graph is that the cavity resonances of the different micropillar cavities consist of distinct troughs. Their number increases as the diameter of the micropillar increases. The figure shows 7, 4, and 2 troughs for micropillars with $20\ \rm{\mu m}$, $6\ \rm{\mu m}$, and $3\ \rm{\mu m}$ diameter, respectively.
To understand these troughs, one should regard the micropillar as a section of a dielectric waveguide bounded by two mirrors \cite{Gerard1996}.
Solving Maxwell's equations with appropriate boundary conditions, one can easily find the fundamental and higher-order guided modes of the waveguide \cite{Snyder1983, Marcuse1991}. 
Since these modes each have a specific propagation constant, their vertical confinement by mirrors result in the formation of a resonant cavity mode at a specific frequency \cite{Gerard1996}.
These troughs in the reflectivity spectrum are due to the coupling of the beam to these resonant modes.

In order to confirm that the reflectivity troughs correspond indeed to the pillar modes, we have compared the reflectivity measurements with room temperature photoluminescence measurements from the embedded quantum dots.
In Fig. \ref{fig:fig3}(b) the measurements for a micropillar with a diameter of $3\ \rm{\mu m}$ are shown.
The spacing between the observed modes is $56\ \rm{cm^{-1}}$ and $55\ \rm{cm^{-1}}$ for the photoluminescence and the reflectivity measurement, respectively, and thus are in very good agreement.
The spectral shift of $12\ \rm{cm^{-1}}$ between the two experiments is attributed to a difference of $\rm{\Delta T\simeq10\ K}$ in temperature between the experiments.

\begin{figure}
\includegraphics[]{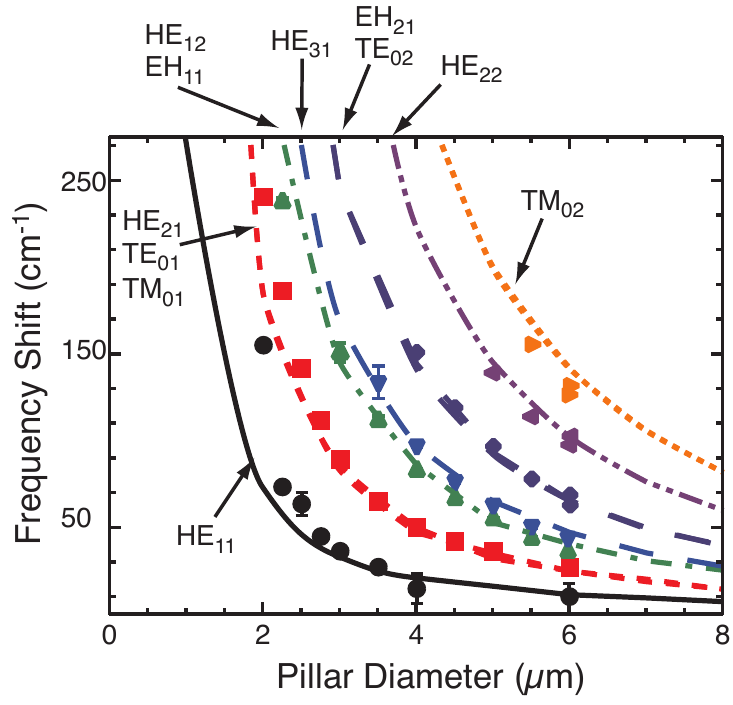}
\caption{\label{fig:fig4}(color online) Measured (symbols) and calculated (curves) mode frequency shift versus pillar diameter. The reference frequency is the cavity resonance of a planar cavity. The agreement between experiment and calculation is very good and shows the mode shift due to the lateral confinement in the micropillar. The modes are labeled using the standard notation for cylindrical dielectric waveguides \cite{Snyder1983, Marcuse1991}.
}
\end{figure}

We have extracted all mode frequencies for the different micropillar diameters from our reflectivity measurements, see Fig \ref{fig:fig3}(a). 
In Fig. \ref{fig:fig4} we have plotted the shift of the mode frequencies vs. pillar diameter with respect to the resonance of a planar cavity. 
The experimentally observed modes are shown as symbols. The calculated frequencies of the modes are plotted as curves. 
The agreement between theory and experiment is strikingly good. The shift of the mode frequency due to the lateral confinement in the micropillar decreases with increasing pillar diameter and converges to an asymptote at $\Delta \omega=0\ \rm{cm^{-1}}$, corresponding to the mode propagation vector $\beta$ in air $\beta_{air}=10650\ \rm{cm^{-1}}$  \cite{Snyder1983}. 
Using a thickness of the $\lambda$-layer of $d=261.7\ \rm{nm}$, as derived from the transfer matrix fit of the micropillar with a diameter of $20\ \rm{\mu m}$, see Fig. \ref{fig:fig1}(c), the calculated mode propagation vector is $\beta=10722\ \rm{cm^{-1}}$. 
The $20\ \mu m$ diameter micropillar was used for the calculation because it is close to the planar case in view of the small beam diameter of approximately $1\ \rm{\mu m}$. 
We can therefore assign the measured modes by their standard notation from waveguide theory \cite{Snyder1983, Marcuse1991}.

\section{Spatial mapping and selective addressing of pillar modes}

We will now discuss how one can couple into all transverse modes in the micropillar with an external incident Gaussian beam. 
We attribute this to the small size of the beam, \textit{i.e.}, being smaller than the pillar diameter.
Early studies of micropillars have experimentally confirmed that a Gaussian beam with a size matched to the pillar diameter can only couple to a few pillar modes \cite{Jewell1989}. 
Our implementation of a small beam, much smaller than the pillar diameter, and the ability to select an off-axis focusing point on the micropillar's top facet lead to a lifting of symmetry constraints and thus allow the coupling to all modes.
Furthermore, the small beam allows us to map the spatial profile of each mode by scanning the beam across the top facet of the micropillar. 

\begin{figure}
\includegraphics[]{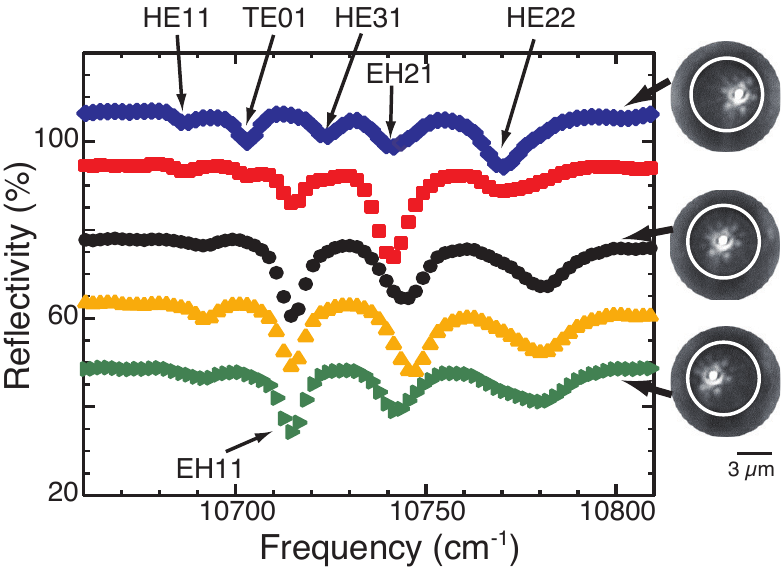}
\caption{\label{fig:fig5}(color online) Reflectivity spectra of the cavity resonance for a $6\ \rm{\mu m}$ diameter micropillar at different lateral positions of the beam. The spectra are offset with respect to each other for a better comparison. The position of the illumination spot is shown exemplarily for three spectra on the microscopy images on the right (the micropillar boundaries denoted by the white circle). The spectrum in the center is measured at the center of the micropillar's top facet.}
\end{figure}

\begin{figure}
\includegraphics[]{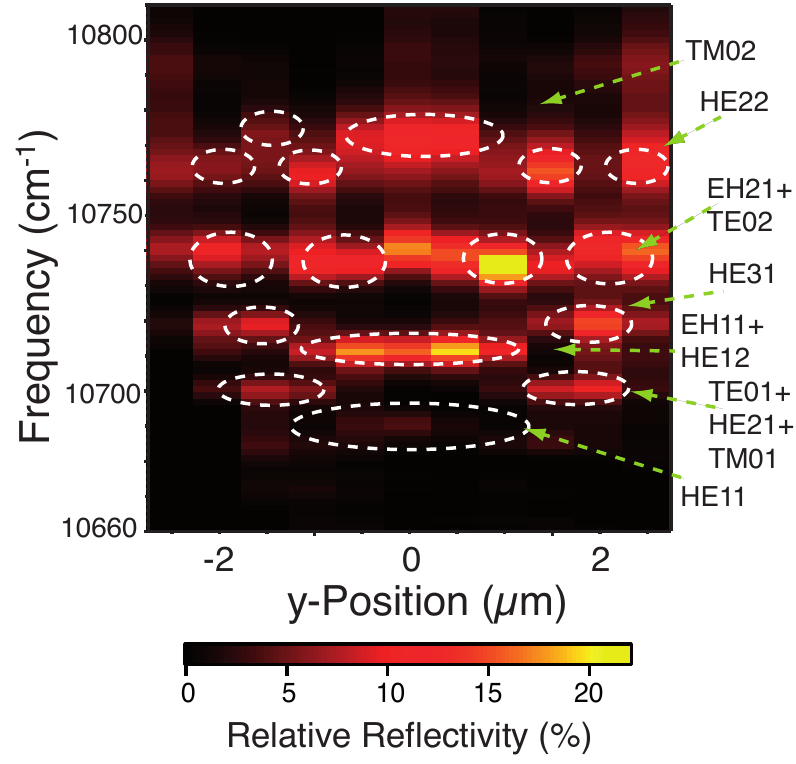}
\caption{\label{fig:fig6}(color online) Colorscale representation of the relative reflectivity at the different transverse positions of the beam vs. the frequency of a $6\ \rm{\mu m}$ diameter micropillar.}
\end{figure}

Figure \ref{fig:fig5} shows reflectivity spectra of the cavity resonance for a $6\ \rm{\mu m}$ diameter micropillar taken at different positions across the equatorial line of the top facet.
Additionally, optical microscopy images, taken with the CCD-camera in the setup, are shown for three selected spectra (top, middle, and bottom). The white circles in the micrographs mark the boundaries of the micropillar. 
There are clear differences visible between the spectra. At each position of the beam the reflectivity of the troughs changes independently. 
To illustrate this spatial dependence of the trough reflectivity we calculated the relative reflectivity intensity of a mode 
\begin{equation}
R^{rel}_{mode}=\frac{R_{mode}-R_{sb}}{R_{sb}},
\end{equation} 
with $R_{sb}$ the stopband reflectivity.
Figure \ref{fig:fig6} shows a colorscale representation of the lateral position-dependent measured relative reflectivity intensity of the $6\ \rm{\mu m}$ diameter micropillar versus frequency. 
The graph can be used as a map to locate the modes in frequency and lateral position. 
One can clearly see that the pillar modes display very different spatial dependences of their trough reflectivity.
As shown below, this behavior directly reflects the different transverse field intensity distributions of the pillar modes. In particular, the position at which the reflectivity trough disappears for a certain mode corresponds to the location of its nodes. As such, these positions strongly depend on the nature of the mode under study. 

\begin{figure}
\includegraphics[]{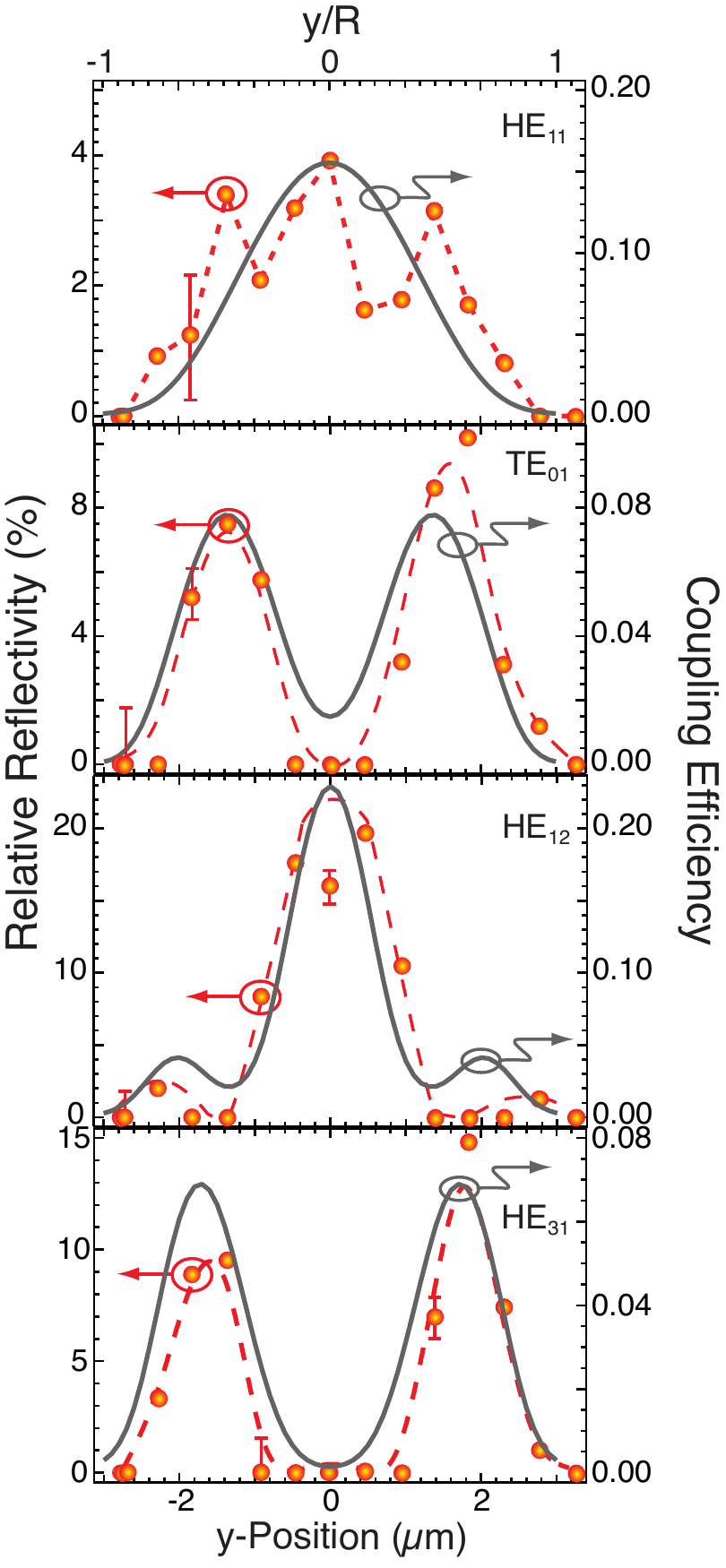}
\caption{\label{fig:fig7}(color online) Cross sections of the measured spatial transverse mode profiles for the first four modes of a $6\ \rm{\mu m}$ diameter micropillar (filled symbols), taken from Fig. \ref{fig:fig6}. The dashed lines are a guide to the eye and typical error bars are included for the measurements. The calculated coupling efficiencies are plotted as solid lines. The calculated line shapes agree very well with the experiments.}
\end{figure}

For a better analysis, cross sections of the first four modes, taken from Fig. \ref{fig:fig6} are shown in Fig. \ref{fig:fig7}. 
The symbols represent the measurements of the relative reflectivity across the top facet of a $6\ \rm{\mu m}$ diameter micropillar. The dashed line connects the data points as a guide to the eye. 
For the $\rm{HE_{11}}$ mode, only one broad maximum is observed since this mode is the fundamental one. Hence this mode has no spatial nodes. 
The energetically next higher mode, the $\rm{TE_{01}}$ mode, shows a node in the center of the micropillar and two anti-nodes at lateral positions $\pm1.5\ \rm{\mu m}$ away from the center. 
For the $\rm{HE_{12}}$ mode we observe a narrow anti-node in the center and two smaller anti-nodes near the edges of the micropillar ($\pm2.3\ \rm{\mu m}$).
The next higher mode, the $\rm{HE_{31}}$ mode, shows again the same symmetry as the $\rm{TE_{01}}$ mode, yet the anti-nodes are now shifted to the brim of the micropillar ($\pm2\ \rm{\mu m}$).
All mode intensities have a symmetry plane at the center due to the cylindrical symmetry of the micropillar.

Up to now we have implicitly assumed that the spatial dependence of the relative reflectivity mimics the mode intensity profile. 
To verify this assumption we have calculated the coupling efficiency of a Gaussian beam to each transverse mode at each position along the equatorial plane of the top facet.
We calculate the coupling efficiency as \cite{Snyder1983}:
\begin{equation}
\label{eq:eq2}
\eta_{lm}=\frac{	\mid \iint dA\ E_{lm} E^*_{in} \mid ^2}{\iint dA\ E_{lm}\cdot E^*_{lm} \cdot \iint dA\ E_{in}\cdot E^*_{in}},
\end{equation}
with $\rm{E_{in}}$ the incident Gaussian field, $\rm{E_{lm}}$ the field of the particular transverse mode, and the asterix denoting the complex conjugate.
Equation \ref{eq:eq2} shows that the coupling of a beam to a specific mode strongly depends on the shape of the incoming field, since it is a measure of the overlap of the electric field profiles of incoming and mode field. The efficiency will therefore even differ for Gaussian incoming beams which have different beam waists coupled to the same cavity mode. The denominator normalizes the coupling to the intensities. 
The calculated coupling efficiencies for each mode are inserted in Fig. \ref{fig:fig7} as solid lines. 
One can see that the calculated line shapes for each mode reflect the symmetry of the mode and reproduce the measurements very well. 
Only the calculated coupling efficiency for $\rm{HE_{11}}$ does not match the experiment. 
We attribute the difference to the actual profile of the incoming beam, which likely differs from the Gaussian beam used in the calculations. This difference is attributed to the assembly of the focussing reflecting objective that shadows the central part of the beam. Thus wavevector components of the beam near $\textrm{k} \approx0$ are blocked. Moreover, the reflecting objective has three radial posts to hold the central mirror that also modify both the incident and reflected beam shapes. 
While a detailed modeling of these beam effects is outside the scope of our paper, the very good quantitative agreement between the calculated coupling efficiencies and the measured intensities for all other modes is gratifying.

\begin{figure}
\includegraphics[]{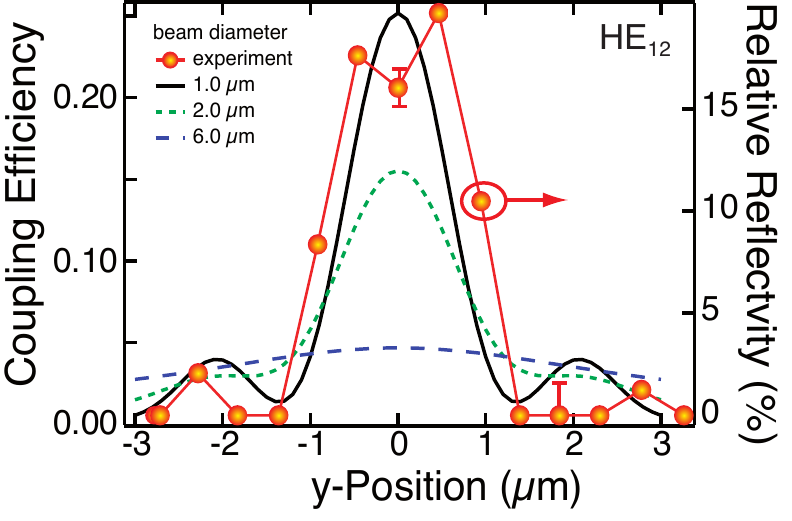}
\caption{\label{fig:fig8}(color online) Calculated coupling efficiencies for different diameter of the incoming beam for the $\rm{HE_{12}}$ mode of the $6\ \rm{\mu m}$ diameter micropillar. With increasing beam diameter the spatial profile of the mode is not visible any more. Additionally, the experiment is shown (filled symbols).}
\end{figure}

The calculations of the coupling efficiencies also explain the experimental fact that we can map the field profiles of the transverse modes with a small-sized beam. 
In Fig. \ref{fig:fig8} we show exemplarily for the $\rm{HE_{12}}$ mode how the coupling efficiency changes as a function of the beam diameter. The measured profile is plotted for comparison (filled symbols and red line).
In the calculations we changed only the beam diameter of the incoming Gaussian beam. With increasing beam diameter two major things become clear, see Fig. \ref{fig:fig8}. First, the coupling efficiency for a focus at the central anti-node decreases. Second, the mode profile cannot be resolved anymore. 
This means that choosing a small beam diameter in the experiment is crucial for resolving the mode profile of the transverse modes. If the beam diameter becomes too big, the mode profiles cannot be resolved anymore and, furthermore, the symmetry constraints apply again reducing the coupling efficiency into all other modes except $\rm{HE_{11}}$.
As a side result of our calculations we can refine our estimate of the beam diameter by matching the calculated mode profile quantitatively to the experimental one, here $\rm{d}\approx1\ \rm{\mu m}$.

Let us finally note that an accurate vertical positioning of the beam is mandatory in this experiment. 
Since we use a high-NA microscope objective to focus our beam onto the top facet of the micropillar, the Rayleigh length is very short. 
Thus the beam waist increases rapidly outside the focus, which would lead to the loss of coupling efficiency into a specific mode.

\section{Prospects}
Having understood why and that in our experiment we are able couple to all transverse optical modes of a micropillar, we return to Fig \ref{fig:fig6}.
It is now clear that we are able to address selectively any resonant mode of a pillar microcavity. By accurately position the beam at a specific location on the micropillar's top facet,  we can allow and favor the coupling to any chosen mode. 
Since the remaining modes, which can be excited at this position, are well separated in frequency, one can ensure that only one mode is excited by choosing the right average frequency and spectral broadening for the beam. For instance $\omega = 10722\ \rm{cm^{-1}}$, $\Delta\omega= 20\ \rm{cm^{-1}}$, and $x=\pm 1.8\ \rm{\mu m}$ leads to solely excitation of the $\rm{HE_{31}}$ mode.
We could use this technique, for example, for the controlled excitation of quantum dots sitting at specific locations in the cavity.
Furthermore, excitation of a single mode in a micropillar using ultrashort pulses can be achieved. 
Since an ultrashort pulse has a corresponding broad spectrum, excitation of a single mode is possible if the bandwidth of the pulse matches the spectral separation of two adjacent modes in the micropillar. The laser bandwidth (and thus the pulse duration) are thus not necessarily limited by the bandwidth of a particular resonance.
For a sub-picosecond Gaussian-shaped pulse $\Delta\tau\leqslant1\ \rm{ps}$ the corresponding bandwidth is $\Delta\omega\geqslant14.67\ \rm{cm^{-1}}$. 
The mode spacing must exceed this value and hence - from Fig. \ref{fig:fig4} - one can derive an upper bound for the micropillar diameter ($d\leqslant5.5\ \rm{\mu m}Ê$). 
Then one has still one extra degree of freedom left namely where to position the beam. This possibility suppresses the excitation of the other possible modes. 
As an example consider the $3\ \mu m$ micropillar shown in Fig. \ref{fig:fig3}(b). It has a mode separation of $55\ \rm{cm^{-1}}$. 
This separation corresponds to a spectral bandwidth of $1.6\ \rm{THz}$ and thus a pulse duration of less than $0.3\ \rm{ps}$ for a Gaussian-shaped pulse. 
With such a pulse duration one would excite both modes. Yet, by positioning the beam at the node position of one mode only the other mode would be excited.
Thus, by matching the pulse width to the mode separation and lateral position the beam on the micropillar, excitation of a single mode with ultrashort pulses is feasible and it is also very attractive in different contexts.
It will allow, for instance, injecting an ultrashort pulse as pump beam into a well-defined mode in optical parametric oscillators based on micropillar cavity polaritons \cite{Bajoni2007}. 
Another possible application domain is the frequency conversion of short light pulses using cavity switching \cite{Preble2007, McCutcheon2007, Tanabe2009}; here again, the selective injection of the pulse into a single discrete pillar mode is highly desirable, so as to get a full control over the light conversion process.   

\section{Conclusion}

We have performed white-light reflectivity measurements of micropillar cavities with diameters ranging from $20\ \rm{\mu m}$ down to $1\ \rm{\mu m}$.
We showed that the use of a beam diameter smaller than the micropillar diameter and the accurate positioning of the beam on the micropillar lifts all symmetry constraints for an efficient coupling of the beam to each transverse mode.
Calculations on the coupling efficiency corroborate our experimental results.
We experimentally resolved and identified the transverse optical modes of the cavity. We furthermore showed that we can map the spatial profile of each mode and that our setup allows us to specifically address a single optical cavity mode. 

\section{Acknowledgement}\label{sec:Acknowledgement}

We thank A. Mosk for valuable discussions and Y. Nowicki-Bringuier for the sample growth.
This work was partly funded through the SMARTMIX Memphis programme of the Netherlands Ministry of Economic Affairs and the Netherlands Ministry of Education, Culture, and Science.
This work is also part of the research programme of the "Stichting voor
Fundamenteel Onderzoek der Materie" (FOM), which is financially supported by the NWO. WLV thanks NWO for a Vici fellowship.

\clearpage


\clearpage

\end{document}